\documentclass[prl,aps, superscriptaddress,twocolumn]{revtex4-1}
\usepackage{amsmath,amssymb}
\usepackage{graphicx}
\usepackage{wasysym}
\usepackage{amsfonts}
\usepackage{bm}
\usepackage{enumerate}
\usepackage{color}
\usepackage[resetlabels]{multibib}
\usepackage{epstopdf}
\usepackage{latexsym}
\usepackage[breaklinks,colorlinks = true,linkcolor = red,urlcolor=cyan,citecolor=red]{hyperref}
\usepackage[caption=false,singlelinecheck=false]{subfig}

\usepackage{times}
\newcommand{\bea}{\begin{eqnarray}}
\newcommand{\eea}{\end{eqnarray}}
\newcommand{\be}{\begin{eqnarray}}
\newcommand{\ee}{\end{eqnarray}}
\newcommand{\bw}{\begin{widetext}}
\newcommand{\ew}{\end{widetext}}

\begin{document}
\title{Topological phases of non-symmorphic crystals : Shastry-Sutherland lattice at integer filling}

\author{Hyeok-Jun Yang}
\email{yang267814@kaist.ac.kr}
\affiliation{Department of Physics, Korea Advanced Institute of Science and Technology, Daejeon, 34141, Korea}
\author{SungBin Lee}
\email{sungbin@kaist.ac.kr}
\affiliation{Department of Physics, Korea Advanced Institute of Science and Technology, Daejeon, 34141, Korea}
\date{\today}
\begin{abstract}
Motivated by intertwined crystal symmetries and topological phases, we study the possible realization of topological insulator in nonsymmorphic crystals at integer fillings. In particular, we consider spin orbit coupled electronic systems of two-dimensional crystal Shastry-Sutherland lattice at integer filling where the gapless line degeneracy is protected by glide reflection symmetry. Based on a simple tight-binding model, we investigate how the topological insulating phase is stabilized by breaking nonsymmorphic symmetries but in the presence of time reversal symmetry and inversion symmetry. In addition, we also discuss the regime where Dirac semimetal is stabilized, having non trivial $Z_2$ invariants even without spin orbit coupling (SOC).
Our study can be extended to more general cases where all lattice symmetries are broken and we also discuss possible application to topological Kondo insulators in nonsymmorphic crystals where crystal symmetries can be spontaneously broken as a function of Kondo coupling. 
\end{abstract}
\maketitle
\label{sec:Introduction}
{\textbf{\textit {  Introduction ---}}}
In the thermodynamic limit, gapped or gapless nature of phases is an important characteristic to classify low energy excitations and their physical properties. For non-interacting system, one can predict so called band insulator where the filling is an integer i.e. unit cell must contain an integer number of electrons per unit cell and spin, thus the band can be completely filled below the Fermi energy. \cite{Faz99} On the other hand, the Mott insulator is a counter example of band insulator where insulating phase with conserving all symmetries are realized even at fractional filling. \cite{mott1974metal, mott1990metal} For such case, the celebrated Hastings-Oshikawa-Lieb-Schultz-Mattis (HOLSM) theorem gives the strong guiding principles for any fractional filling no matter what types of particles and interaction strength; If the system at fractional filling do preserve all the symmetries, it must be either gapless or gapped with degenerate ground states that accompany fractional low energy excitations. \cite{lieb1961two, oshikawa2000commensurability, hastings2004lieb, hastings2005sufficient,essin2014spectroscopic}

%
In crystals, it turns out that discrete lattice symmetries can give similar constraints even at integer fillings. \cite{parameswaran2013topological, konig1997electronic,konig2000symmetry,watanabe2015filling,watanabe2016filling} In particular, it holds for nonsymmorphic crystals where their space group symmetries are not represented by a direct product of translation and point group symmetry, thus always contain glide reflections or screw rotations. These symmetries accompany fractional (say $1/\mathcal{S}$) translation followed by either reflection or rotational symmetries. Attributed to such fractional translation, the filling $\nu$ to be a trivial insulator is typically multiple of specific integer $\mathcal{S}$, i.e. $\nu \!=\! n \mathcal{S}$, $n\in Z$. Here, we emphasize the filling $\nu$ is defined as the average number of electrons in a unit cell for each spin polarization. For any other integer fillings ($\nu \! \not \in \! n \mathcal{S}$), one can still apply HOLSM theorem and the system with preserving all the symmetries must be categorized into two cases; (i) gapless (ii) gapped with fractional low energy excitations. \cite{lee2016fractionalizing} 
This strong argument indicates if we ignore exotic scenario of Mott insulating phases with fractional low energy excitations, then the gaplessness of the system is protected by nonsymmorphic crystal symmetries at certain integer fillings $\nu \! \not \in \! n \mathcal{S}$.

One of the intriguing question is then how the system drive into the transition from gapless semimetal to gapped insulating phase by breaking nonsymmorphic crystal symmetries. In particular, when the electronic system is described by heavy ions, the SOC effect play an important role and it is natural to consider the interplay of nonsymmorphic symmetry breaking and SOC results in unique topological insulating phases.  
Such gapped phases can be generally favored to reduce the kinetic energy of electrons, thus it gives rise to the instability of gapless semimetallic phase protected by either glide reflection symmetry or screw rotation symmetry at integer filling. Therefore, it could drive the system into an insulating phase with spontaneous breaking of nonsymmorphic crystal symmetries, accompanied with lattice distortion, formation of charge (spin) ordering. \cite{coleman2006heavy}
Furthermore, one can also expect our scenario applicable to the Kondo lattice system in nonsymmorphic crystals. \cite{chang2017mobius,dzero2010topological} When  localized magnetic moments and itinerant electrons are both present, the control of Kondo coupling strength can derive multiple phase transitions. At  particular integer filling of itinerant electrons, an intermediate Kondo coupling leads to partial Kondo screening in such a way that all lattice symmetries are broken. Then the system could be driven into the {\it topological Kondo insulator} with spotaneously broken nonsymmorphicity, which will be a natural extension of earlier work studied in Ref.\onlinecite{pixley2016filling}. 
%


In this paper, we investigate possible phase transition from a gapless semimetal protected by nonsymmorphic crystal symmetry to a topological insulator where the crystal symmetries are broken but edge states are protected by time reversal symmetry. \cite{hasan2010colloquium, qi2011topological, bernevig2013topological} 
 We exemplify our scenario to spin-orbit coupled electronic system in a specific two dimensional crystal, Shastry-Sutherland lattice (SSL). \cite{shastry1981exact} Especially, we focus on the filling $\nu\!=\!1$ per unit cell and spin where the gapless electronic structure at the Fermi level is protected by glide reflection symmetry. Considering two different ways of breaking glide reflection symmetry, the  stabilities of trivial insulator and topological insulator are addressed as functions of electron hopping and strength of SOC. \cite{young2015dirac, kane2005z}
Based on the calculation of $Z_2$ invariants \cite{fu2007topological, hughes2011inversion}, we show large parameter space where the topological insulating phase is indeed stabilized. Furthermore, we discuss the parameter space where the odd number of gapless Dirac points must be present in the absence of SOC.  
We note that similar argument can be easily extended to other nonsymmorphic crystals including both two dimensional and three dimensional lattices.

\label{sec:Extended HOLSM theorem}
{\textbf{\textit { Review --- Extended HOLSM theorem in nonsymmrophic crystals :}}}
We first briefly review the HOLSM theorem to study the necessary condition for trivial insulators in the presence of U(1) symmetry. Here, the U(1) symmetry corresponds to the fixed number of electrons (or magnetization plateau) in electronic system (or spin system) respectively. According to the HOLSM theorem, an incommensurability is a robust mechanism to protect gapless states. Suppose the three dimensional system is gapped and periodic along each direction having lengths $L_x, L_y, L_z$ respectively so there are $V\!=\!L_x \!\times\! L_y \!\times \! L_z$ number of unit cell. With the total particle number $N$ in the system, the filling can be defined as {$\nu\!=\!N/sV$ where $s$ counts the spin degeneracy if any}. The HOLSM theorem states that at fractional filling $\nu$, a trivial insulator which respects all lattice symmetries with unique ground state is forbidden. For gapped phases, only two alternatives are allowed;  the system must break translational symmetry to enlarge the unit cell and restore the commensurability, or it preserves all symmetries but develops a topological order described by ground state degeneracy. \cite{lieb1961two, oshikawa2000commensurability, hastings2004lieb}

In order to sketch this theorem, Laughlin's argument is applied. {\cite{laughlin1981quantized, oshikawa2000commensurability}} By threading a flux quantum encircled by the periodic direction, the Hamiltonian evolves into another gauge-equivalent Hamiltonian with the new ground state.
If the flux threading is in adiabatic process, the new Hamiltonian after flux insertion should go back to the original Hamiltonian through a proper gauge transformation but the ground state may not. Whether this new ground state is the same as the original one, can be investigated by comparing their quantum numbers. 
When the system preserves the translational symmetry, the Hamiltonian ${\mathcal{H}}$ and the translation operator $T_x$ commute with each other $[{\mathcal{H}},T_x] \!=\!0$ and the ground state $|\Psi\rangle$ can be chosen to have a definite crystal momentum $P_{x}$ where $T_x\!=\!e^{iP_x}$. By threading a flux quantum $2\pi$ in the units of $\hslash \!=\!c\!=\!e\!=\!1$ adiabatically, there is no Aharonov-Bohm effect and the resulitng Hamiltonian ${\mathcal{H}}(2\pi)$ is gauge-equivalent to the original one ${\mathcal{H}}(0)$. This is acheived by adding an uniform vector potential \textit{\textbf{A}}$=(2\pi/L_x)\hat{x}$. During the flux insertion, the translational symmetry is maintained thus the momentum of the new eigenstate $|\Psi'\rangle$ is not changed. Now the large gauge transformation ${U}_x$ is applied to the new Hamiltonian $\mathcal{H}(2\pi)$, defined as ${U}_x\!=\! \text{exp}[2\pi i{\int d^d {\boldsymbol r}{~ r_x{\rho}({\boldsymbol r})/L_x}}]$ where $\rho ({\boldsymbol r})$ and $r_x$ are the particle density and the $x$ component at position ${\boldsymbol r}$ respectively. Then the original Hamiltonian is restored as $U_x {\mathcal H}(2\pi) U_x^{\dagger}\!=\!{\mathcal H}(0)$ and the eigenstate evolves as $|\Psi'\rangle \!\rightarrow \! U_x|{\Psi'}\rangle \!\equiv \! |\widetilde{\Psi}\rangle$. \cite{oshikawa2000commensurability, hastings2005sufficient}  Here, the non-commutativity of ${U}_x$ with $P_x$ leads to the non-trivial results.
\bea
{U_x}^{\dagger}T_x{U_x}\!=\!T_x\textup{exp}[2\pi i\frac{N}{L_x}]=T_x\textup{exp}[2\pi i \nu C],
\label{eq:HOLSM}
\eea
where $\nu$ is the filling and $C$ is the cross-sectional area. Then the momentum changes after the flux insertion as $P_x \!\rightarrow \! P_x\!+\!2\pi \nu C$ followed by gauge transformation.  This can be also understood by the Faraday's law for charge particles. The flux insertion during $\Delta T$ induces an opposing electric field $E$. Each charge experiences electromotive forces as {\cite{parameswaran2013topological}},
\bea
E\cdot L_x=\frac{d}{dt}\int_0^{L_x}A_x dx=\frac{2\pi}{\Delta T}
\label{eq:EM force}
\eea
Thus the momentum of $N$-particle system changes $\Delta P_x=NE\Delta T=2\pi N/L_x$ during $\Delta T$.
The distinct crystal momentum after the insertion implies the topological degeneracy. At fractional filling $\nu\!=\!p/q$ ($p$ and $q$ are coprime integers), the changed momentum is distinct from the original one (assuming $C$ is coprime to $q$), which implies the topological order with at least $q$ degeneracy. Otherwise the system needs to break translational symmetry to enlarge the unit cell.

 At integer filling $\nu \!\in \!Z$, the crystal momentum fails to differentiate the distinct eigenstates. However, referring the expression {Eq. \eqref{eq:HOLSM}}, it is expected that the quantum numbers of a lattice symmetry operator containing fractional translation may inherit the role at integer filling. Non-symmorphic symmetries play such a role and enable the system at integer filling to differentiate between the $|\Psi\rangle$ and $|\widetilde{\Psi}\rangle$. \cite{parameswaran2013topological, konig1997electronic,konig2000symmetry} Consider a non-symmorphic symmetry operator $G_x$ which can be represented as $G_x\!=\!g*(T_{x/S_G})$ where $g$ is a point group element and $T_{x/S_G}$ is a $1/S_G$ fractional translation along $\hat{x}$-direction. Here, $\hat{x}$-direction is chosen to be invariant under $g$. For instance, the glide reflection symmetries corresponds to $g$ to be mirror reflection and $S_G\!=\!2$, while the $n$-fold screw rotation symmetry corresponds to $g$ to be rotation and $S_G\!=\!n$. 

For non-symmorphic crystals, the Hamiltonian commutes with $G_x$, $[\mathcal{H},G_x]\!=\!0$ and the ground state can be chosen to be an eigenstate of $G_x$. After the flux insertion followed by the gauge transformation $U_x$ as before, one can easily check that the quantum number of $G_x$ is changed as 
\bea
{U_x}^{\dagger}G_x{U_x}\!=\!G_x\textup{exp}[2\pi i \frac{\nu C}{S_G}],
\label{eq:extended HOLSM}
\eea
Again, assuming that $C$ is coprime to both $q$ and $S_G$, filling $\nu\!=\!p/q$, the topological degeneracy increases $S_G$ times. In conclusion, the gapped state requires either at least $q  S_G$ topological degeneracy, or the filling must be a multiple integer of $S_G$. Otherwise, the system is gapless unless lattice symmetries are broken. 
This argument holds for any dimensions and particle species. The utility of this theorem is that it holds even in the presence of interactions and does not rely on perturbative approaches. In bulk band dispersions, $S_G$ detached bands in non-symmorphic crystals are the evidence of this feature. \cite{young2015dirac, michel1999connectivity, topp2016non, ekahana2017observation} \\

\begin{figure}[t]
\centering
\subfloat[]{\label{fig:SSL}\includegraphics[width=0.532\columnwidth]{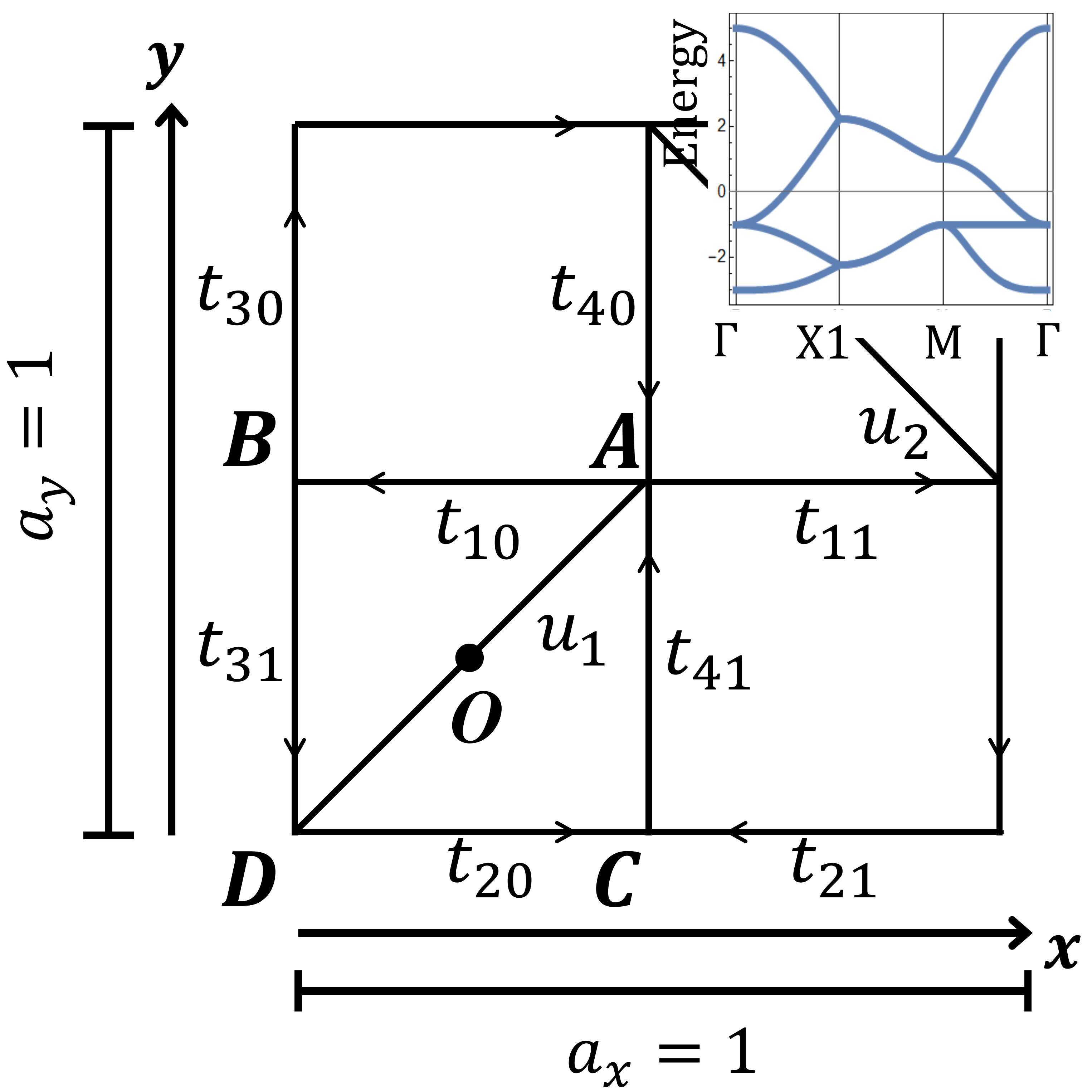}}
\subfloat[]{\label{fig:M point}\includegraphics[width=0.44\columnwidth]{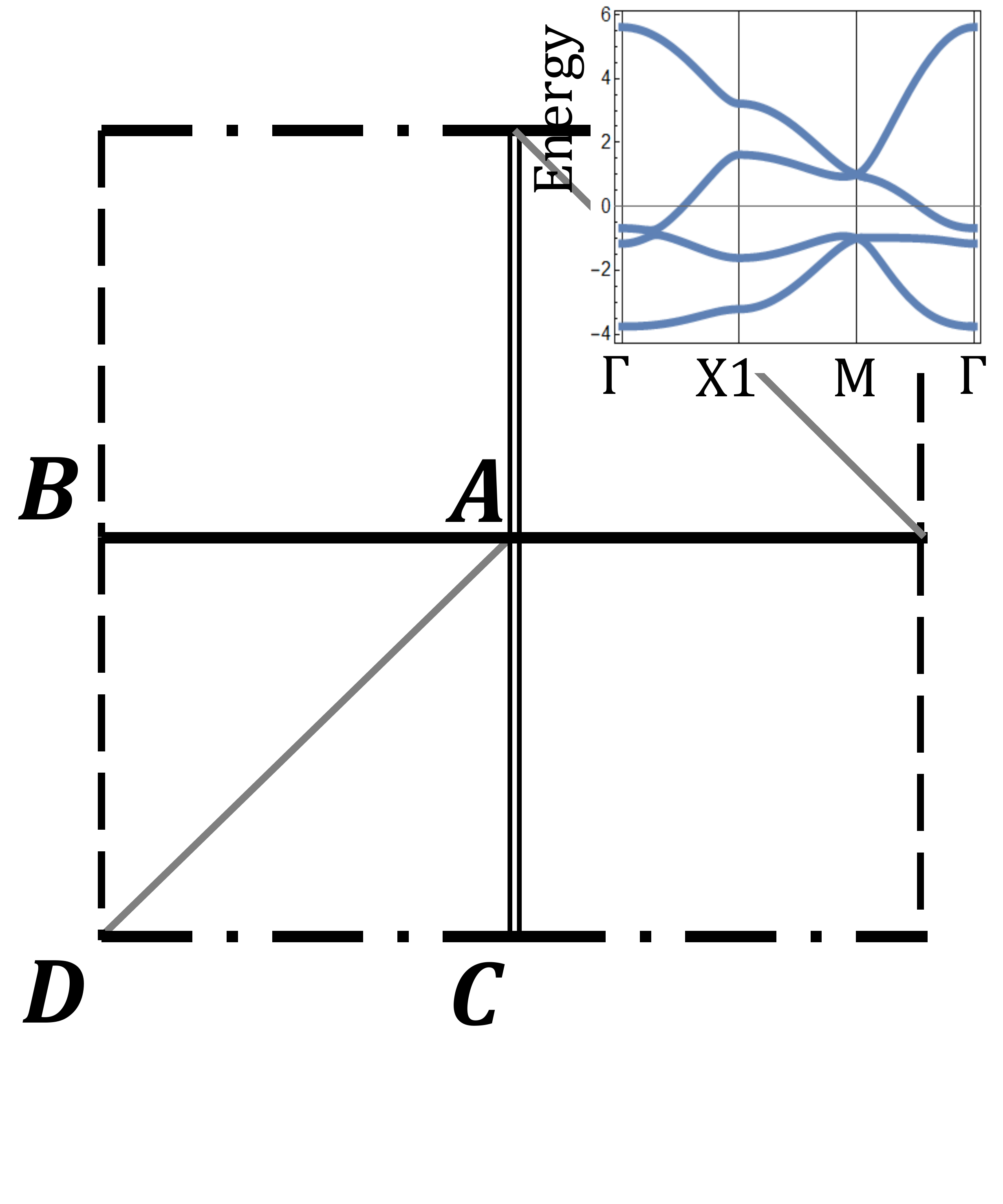}}
\caption{(a) Shastry-Sutherland lattice with nearest-neighbor hopping parameters $t_{ij}$ and $u_i$. The arrow on each link indicates the direction of symmetry allowed SOC. (Inset) bulk dispersion when all parameters are equal to 1. Along the line X1-M, the first and second bands (the third and fourth bands) are degenerate and these degeneracies are protected by glide reflection symmetry. (b) In Shastry-Sutherland lattice, solid, dashed, dashdotted and double lines indicate distinct electron hopping configurations which break all lattice symmetries. (Inset) bulk dispersion with all distinct hoppings described by different line styles. The line degeneracy along X1-M is absent but  the degeneracy at M point is still present.}
\end{figure}

\label{sec:Band properties in nonsymmorphic SSL}
{\textit{\textbf{ Band properties in nonsymmorphic SSL ---}}}
Focusing on electronic band structure of nonsymmorphic crystals, we explore how the nonsymmorphic symmetry breaking leads to non-trivial topological insulating phases. In particular, we employ the simple tight binding model on SSL (space group $p4g$) with nearest neighbor hoppings including horizontal ($-$), vertical ($\mid$) and diagonal ($\diagup$) directions. 
\bea
H_0 \!=\! \sum_{\sigma, \langle i,j \rangle \in \mid , -}t_{ij} c^\dagger_{i \sigma} c_{j \sigma} \!+\!\sum_{\sigma, \langle i,j \rangle \in \diagup}u_{ij} c^\dagger_{i \sigma} c_{j \sigma},
\label{eqn:tight-binding}
\eea
where $c^{(\dagger)}_{i,\sigma}$ indicates (creation) annihilation operator at site $i$ and spin $\sigma$. Each hopping parameter $t_{ij}$ and $u_{ij}$ is depicted in {Fig. {\ref{fig:SSL}}}.
For simplicity, all $t_{ij}$ and $u_{ij}$ are chosen to be positive and spin-independent. From now on, we particularly focus on the filling $\nu\!=\!1$ per unit cell and spin but without losing generality similar argument can be done for another case $\nu\!=\!3$. 

First, let's consider the spinless system and 
each unit cell contains one electron on average, thus filling $\nu\!=\!1$. The unit cell of SSL contains 4 sites and the lattice symmetries such as inversion $P$, 
$C_4$ rotation and mirror symmetries $M_{\hat{x}+\hat{y}}, M_{-\hat{x}+\hat{y}}$ are present when hopping parameters $t_{ij}$ are all the same and $u_1\!=\!u_2$. (See Fig. \ref{fig:SSL}.) \cite{shastry1981exact, sriram2002srcu2} In addition, the glide reflection symmetries $G_x$, $G_y$ defined by half-translations along $x,y$-direction followed by mirror reflections $M_{\hat{y}}, M_{\hat{x}}$, are present. 
Inset of Fig. \ref{fig:SSL} shows the band structure along the lines with high symmetry points $\Gamma(0,0)$, X1$(\pi,0)$ (X2$(0,\pi)$) and M$(\pi,\pi)$. 
At $\nu\!=\!2$, the system is gapless at $\Gamma$ and becomes a semimetal. \cite{kariyado2013symmetry}
For $\nu\!=\!1$ and $3$, one can see that the first and second bands (the third and fourth bands) are degenerate along the lines X1-M and X2-M. 
These line degeneracies are protected by glide reflection symmetries, $G_x$ and $G_y$ respectively. In the absence of spins, $G_x^2\!=\!e^{ik_x}\!=\!-1$ along the Brillouin zone boundary $k_x\!=\!\pi$. Then there exists additional degeneracy related by $[h_0(\pi, k_y),G_x\Theta]\!=\!0$ and $(G_x\Theta)^2\!=\!-1$ where $h_0(k_x,k_y)$ is the Hamiltonian matrix for Eq. \eqref{eqn:tight-binding} at a momentum $(k_x,k_y)$ and $\Theta\!=\!K$ is the complex-conjugate operator. 
Similarly, the invariant line $k_y=\pi$ has a line degeneracy protected by $G_y\Theta$. 

 The insulating phases at $\nu\!=\!1$ and $3$ thus always require breaking of $G_x$ and $G_y$ symmetries, but interestingly within the simple tight binding model, the inverse is not always true. Suppose a system in which electrons with $s$-orbitals sit on SSL at integer filling. The spin degrees of freedom is not considered yet. The Hamiltonian in momentum space can be rewritten as ${H}_0=\Sigma_{\textbf{k}}\psi_{\textbf{k}}^{\dagger} h_0(\textbf{k})\psi_{\textbf{k}}$ where $\psi_{\textbf{k}}=(c_{{\textbf{k}},A},c_{{\textbf{k}},B}, c_{{\textbf{k}},C},c_{{\textbf{k}},D})^{T}$ and $c_{\textbf{k},\alpha}^{(\dagger)}$ is annihilation (creation) operator at momentum $\textbf{k}$ and sublattice $\alpha$. 
When $t_{ij}$ are all equivalent and $u_1=u_2$, the Hamiltonian at M point can be readily analyzed. 
The eigenspace at M point is spanned by $|\Psi1_\pm \rangle \!=\!\frac{1}{\sqrt[]{2}}(c_{{M},A}^{\dagger}\! \pm \!c_{{M},D}^{\dagger})|0\rangle$ and $|\Psi2_\pm \rangle \!=\!\frac{1}{\sqrt[]{2}}(c_{{M},B}^{\dagger} \! \pm \!c_{{M},C}^{\dagger})|0\rangle$ with eigenvalues $\pm u_1$ for occupied (-) and unoccupied (+) eigenspace respectively. In order to open a gap, these states need to be coupled through the Hamiltonian, e.g. $\langle \Psi2_a|h_0(M)|\Psi1_a\rangle\!\neq\!0$. One may think breaking lattice symmetries by varying hopping parameters $t_{ij}$ and $u_i$ leads to finite couplings between $|\Psi1_a \rangle$ and $|\Psi2_a \rangle$ thus to open a gap in the system.
 However, there exists the case where breaking of all lattice symmetries still makes their coupling to be zero.  In {Fig. {\ref{fig:M point}}}, the links sketched with the same styles indicate the identical hopping parameters along the links, i.e. $t_{10}\!=\!t_{11}$, $t_{20}\!=\!t_{21}$, $t_{30}\!=\!t_{31}$, $t_{40}\!=\!t_{41}$.   This configuration breaks all lattice symmetries but the degeneracy at M point is preserved. (See the band structure in the inset of Fig. \ref{fig:M point}.) The reason is as following. In $\textbf{k}$-space, the electron hoppings between sublattices are performed by consuming a phase of the wavepacket. At M point, the phases from one site to another sublattices along $x,y$ directions are cooperatively canceled thus their off-diagonal components of $h_0(M)$ vanish, resulting in degenerate bands at M point. Of course, the different hopping parameters between $\hat{x}/2$ and $-\hat{x}/2$ directions (similarly for $\hat{y}$ direction) will lead the system to be gapped which are discussed below. 

Now let's consider the spinful electronic system. Including spin degeneracy, there exist eight bands in total and each two bands are degenerate since Kramers doublet $(P\Theta)^{2}\!=\!-1$. Thus, at fillings $\nu\!=\!1$ and $\nu\!=\!3$ (per unit cell and spin), four bands are degenerate along X1-M and X2-M as shown in Fig. \ref{fig:SSL}. (Two from Kramers doublet and two from glide reflection symmetries.)
The intrinsic spin orbit coupling can be included as an imaginary hopping term as following.
\bea
H_{SO} = \sum_{\langle i,j \rangle, \sigma}i \sigma_{z} d_{ij} c^\dagger_{i \sigma} c_{j \sigma},
\label{eq:SOC}
\eea
where $d_{ij}$ indicates the SOC strength. 
In the presence of intrinsic SOC, the total spin $S_z$ is still conserved so one can consider the system for each spin independently. 
For each spin sector, this imaginary hopping can be considered as an effective magnetic flux which is required to obtain non-trivial insulating phases. \cite{haldane1988model, kane2005z, kane2005quantum} With preserving all symmetries in SSL, the magnitude of parameters $d_{ij}$ should be equivalent say $|d_{ij}|\!=\!\lambda$ and their relative $\pm$ signs are depicted in Fig. \ref{fig:SSL}, i.e. the arrow toward $i$ from $j$ defines $d_{ij}\!=\!+\lambda$.
Such SOC term opens a gap at half filling $\nu\!=\!2$ and leads to a topological insulating phase. In the same manner as discussed in graphene at half filling, the topological insulator is stabilized by considering SOC. \cite{kane2005z, kane2005quantum} In this case, the filling is a multiple integer of fractional translation $\mathcal{S}\!=\!2$, thus the HOLSM theorem is silent and the system can be either gapless or gapped depending on controlling parameters.\cite{parameswaran2013topological, kariyado2014emergence} Indeed, the presence and absence of SOC makes the system either gapless or gapped at this filling. 

\begin{figure}[t]
\subfloat[]{\label{fig:distort1}\includegraphics[width=0.23\textwidth]{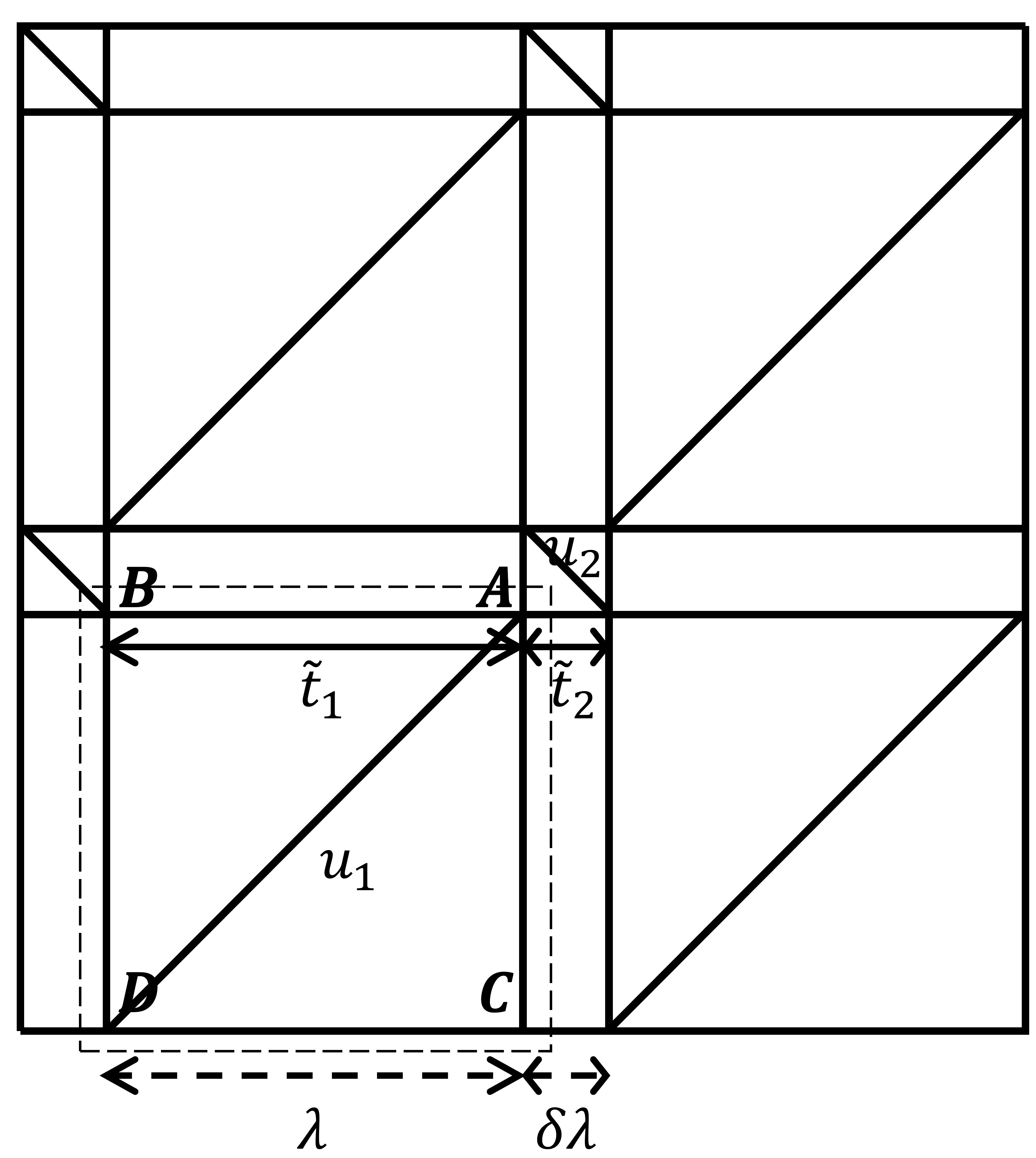}}
\subfloat[]{\label{fig:distort2}\includegraphics[width=0.23\textwidth]{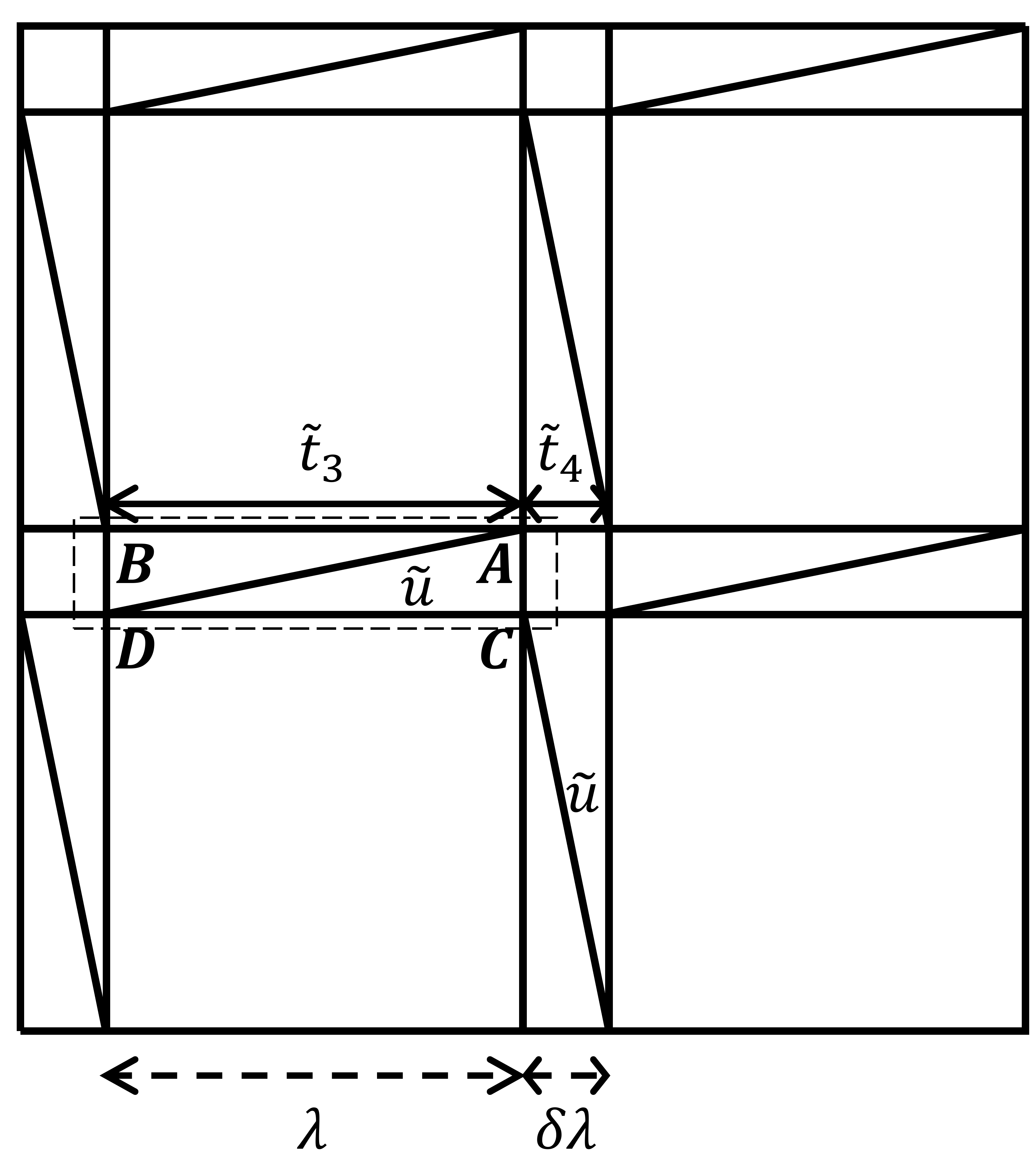}}
\caption{Distorted Shastry-Sutherland lattice with broken glide reflection symmetries but preserving mirror or rotation symmetries; (a) Preserving mirror symmetries with hopping parameters defined as Eq. \eqref{eq:param} (i), (b) preserving $C_4$ rotation symmetry with hopping parameters defined as Eq. \eqref{eq:param} (ii). Dashed boxes indicate the unit cell which includes four sublattices $A$,$B$,$C$ and $D$. $\delta$ measures the relative distortion in SOC. By parameterizing $\tilde{t}_i$, $\lambda$ and $\delta$, Fig. \ref{fig:phase} shows the region where topological insulating phases are stabilized. See the main text for more details.}
\label{fig:distort}
\end{figure}

\label{sec: Broken glide reflection symmetries and topological phases}
{\textit{\textbf{ Broken glide reflection symmetries and topological phases ---}}}
At filling $\nu\!=\!1$ and 3, the band degeneracies along X1-M and X2-M are protected by glide reflection symmetry. \cite{young2015dirac}
To explore how band degeneracies split and the system goes into an insulating phase, we can consider two particular cases that break glide reflection symmetries illustrated in Fig. \ref{fig:distort}. 
\begin{eqnarray}
\label{eq:param}
\text{(i) Fig. \ref{fig:distort1}} ~- ~&& \tilde{t}_1 \! \equiv \! t_{10}\!=\!t_{20}\!=\!t_{31}\!=\!t_{41},  \\
&& \tilde{t}_2\! \equiv \! t_{11}\!=\!t_{21}\!=\!t_{30}\!=\!t_{40}. \nonumber \\
~\text{(ii) Fig. \ref{fig:distort2}}~ -~ && \tilde{t}_3 \! \equiv \! t_{10}\!=\!t_{20}\!=\!t_{30}\!=\!t_{40},\nonumber \\
&& \tilde{t}_4\!\equiv\!t_{11}\!=\!t_{21}\!=\!t_{31}\!=\!t_{41},
\tilde{u}\!\equiv\!u_1\!=\!u_2. \nonumber
\end{eqnarray}
For case (i) in Eq. \eqref{eq:param}, the system preserves two mirror reflections $M_{{\hat{x}\pm\hat{y}}}$ but breaks all other symmetries $G_x, G_y, C_4$, whereas in case (ii), the system preserves $C_4$ but breaks $G_x, G_y, M_{{\hat{x}\pm\hat{y}}}$. In both cases, glide symmetries are broken. 
The SOC strength $d_{ij}$ is also modified based on the broken spatial symmetries of each case (i) and (ii).  
As depicted in Figs. \ref{fig:distort1} and \ref{fig:distort2}, the deviation of SOC is represented as the ratio of two adjacent strengths $\delta$. When $\delta\!=\!1$, the SOC term recovers full lattice symmetries of original Shastry-Sutherland lattice. 
\begin{figure}[t]
\subfloat[Fig. \ref{fig:distort1} -- ${u_1}\!=\!2$, ${u_2}\!=\!4$, $\delta\!=\!-2$]{\label{fig:phase1}\includegraphics[width=0.235\textwidth]{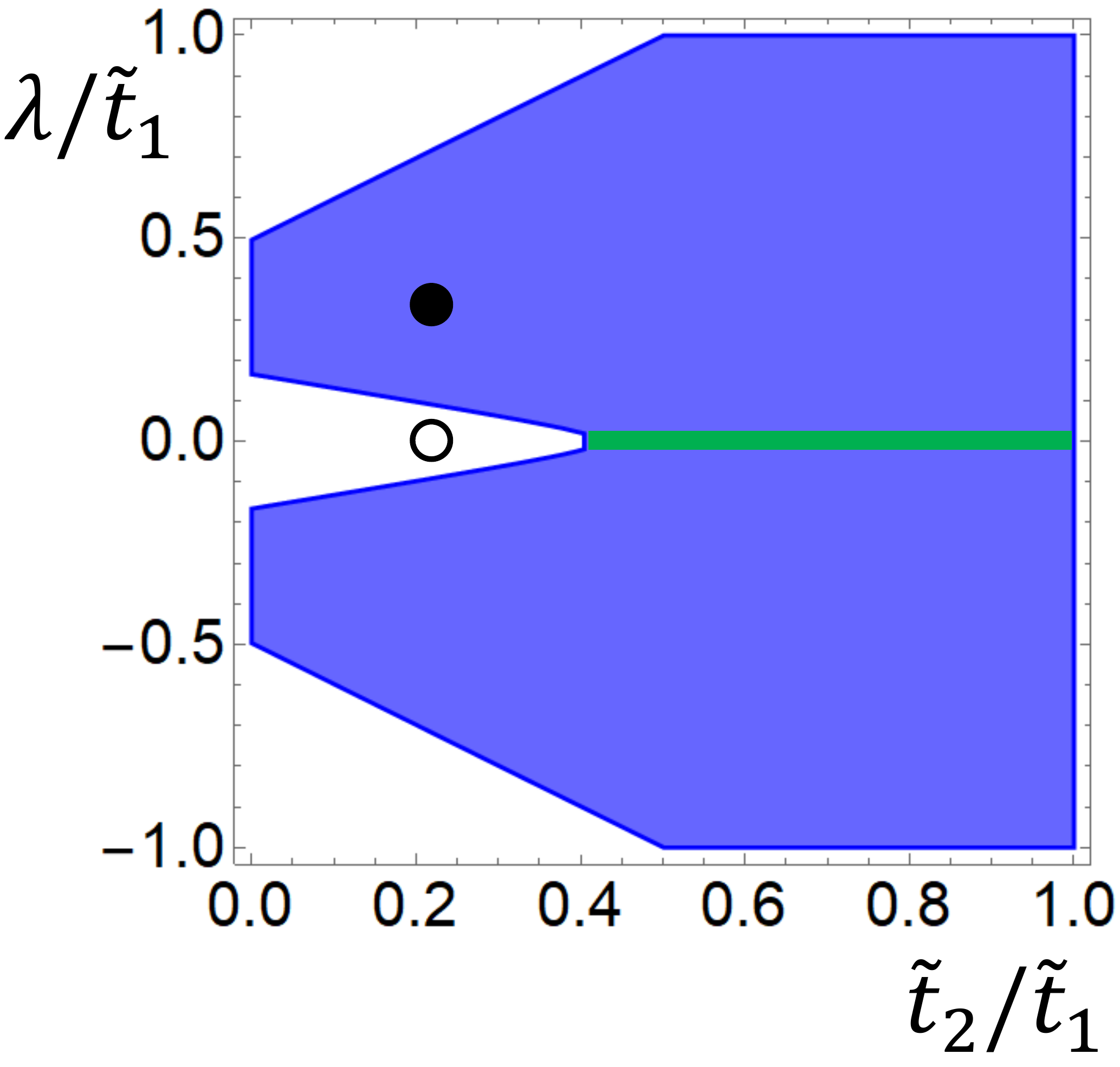}}
\subfloat[Fig. \ref{fig:distort1} -- ${u_1}\!=\!2$, ${u_2}\!=\!4$, $\delta\!=\!2$]{\label{fig:phase2}\includegraphics[width=0.235\textwidth]{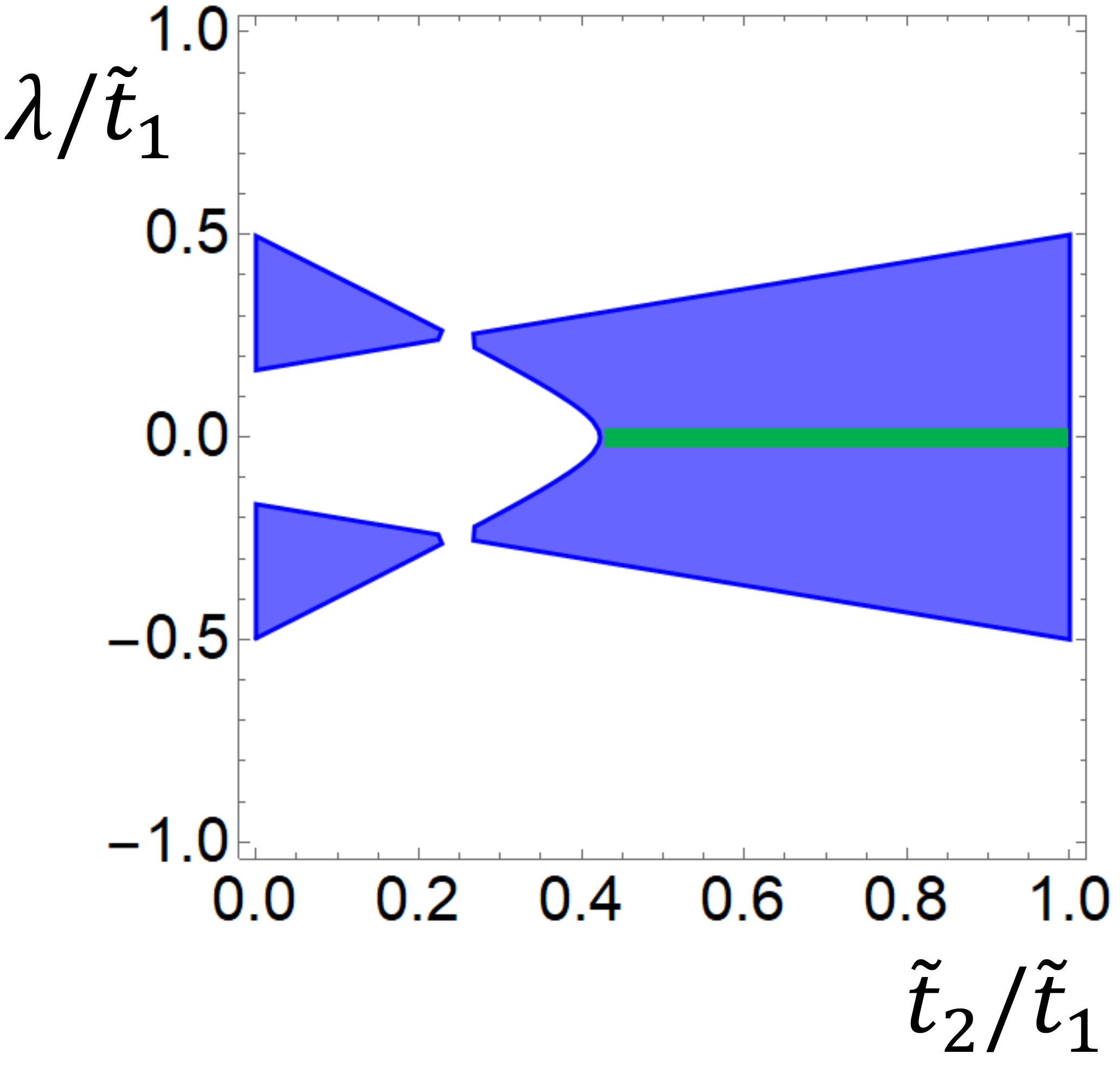}}
\\
\subfloat[Fig. \ref{fig:distort2} -- $\tilde{u}\!=\!1$, $\delta \!=\!-2$]{\label{fig:phase3}\includegraphics[width=0.235\textwidth]{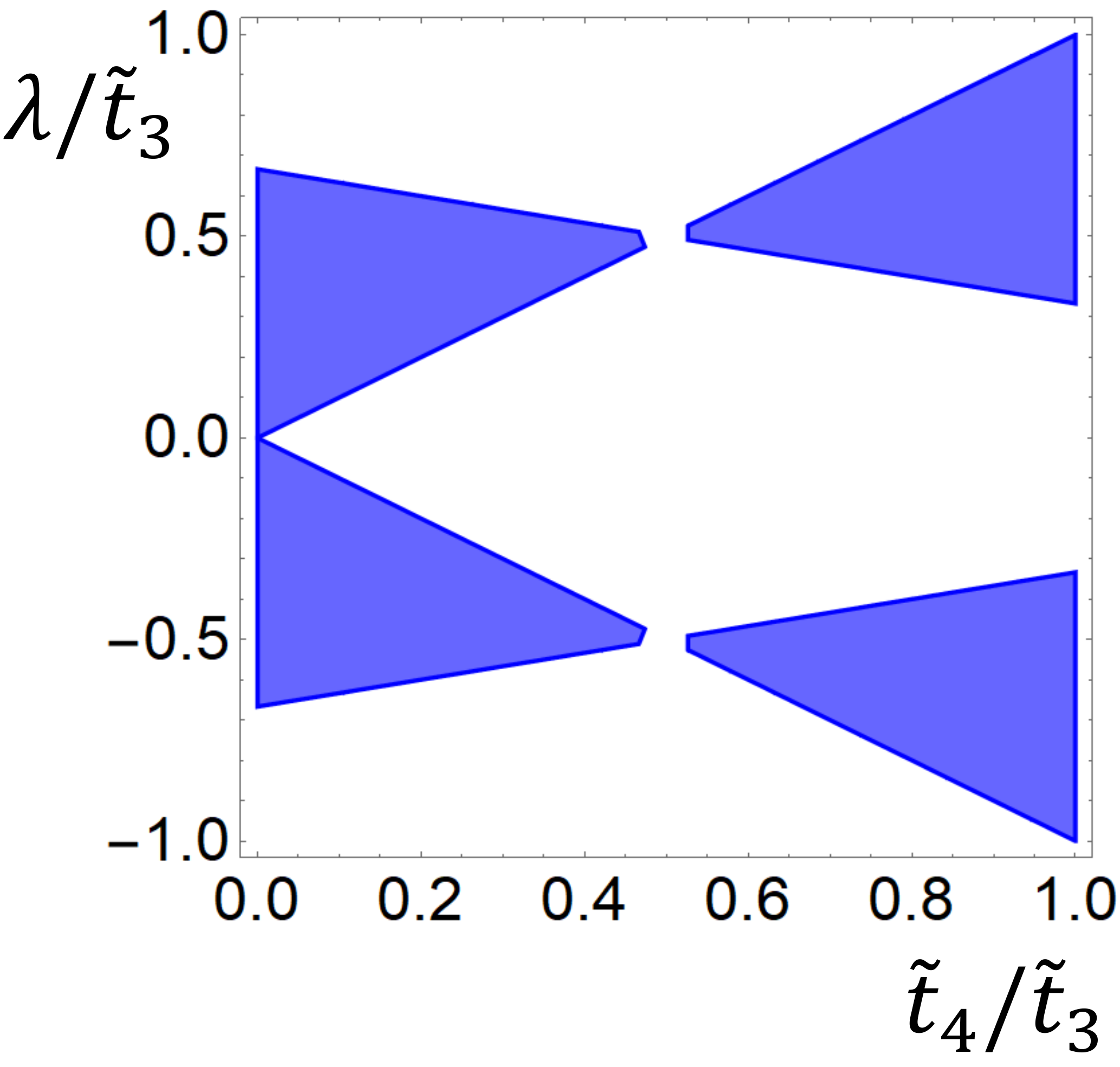}}
\subfloat[Fig. \ref{fig:distort2} -- $\tilde{u}\!=\!1$, $\delta \!=\!2$]{\label{fig:phase4}\includegraphics[width=0.235\textwidth]{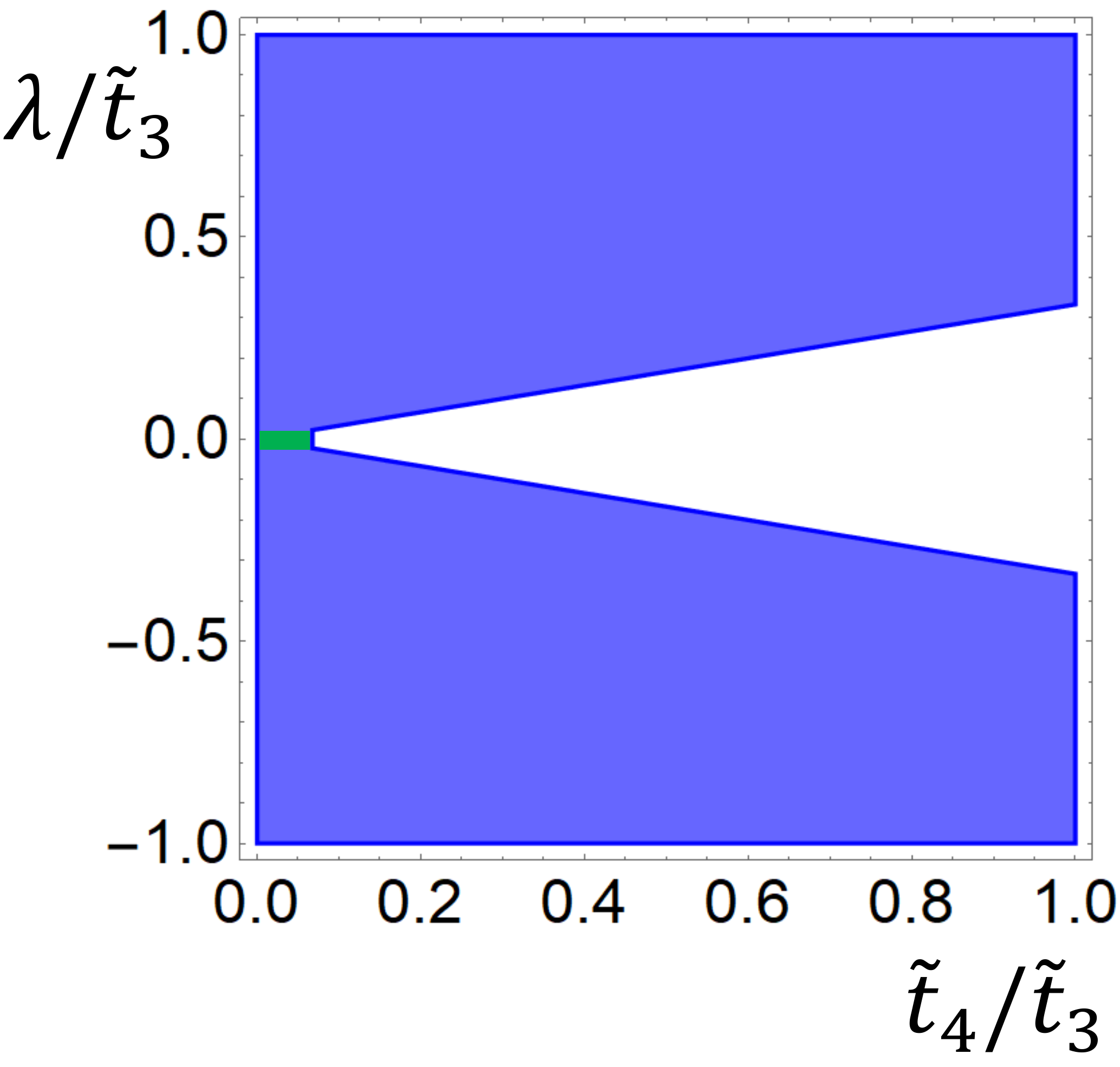}}
\caption{(color online) Phase diagrams of trivial and topological insulators for gapped systems, as functions of $\tilde{t}_2/\tilde{t}_1$ and $\lambda$/$\tilde{t}_{1(3)}$; (a) and (b) are for the distorted lattice structure illustrated in Fig. \ref{fig:distort1} and (c) and (d) are for the cases illustrated in Fig. \ref{fig:distort2}. Blue and white color regime represent where topological insulator and trivial insulator are stabilized respectively.
Green colored region at $\lambda=0$ is where $Z_2$ invariant is -1 and Dirac points or Fermi arc enclosing Dirac points are stabilized.} 
\label{fig:phase}
\end{figure}
\begin{figure}[t]
\subfloat[]{\label{fig:trivial}\includegraphics[width=0.169\textwidth]{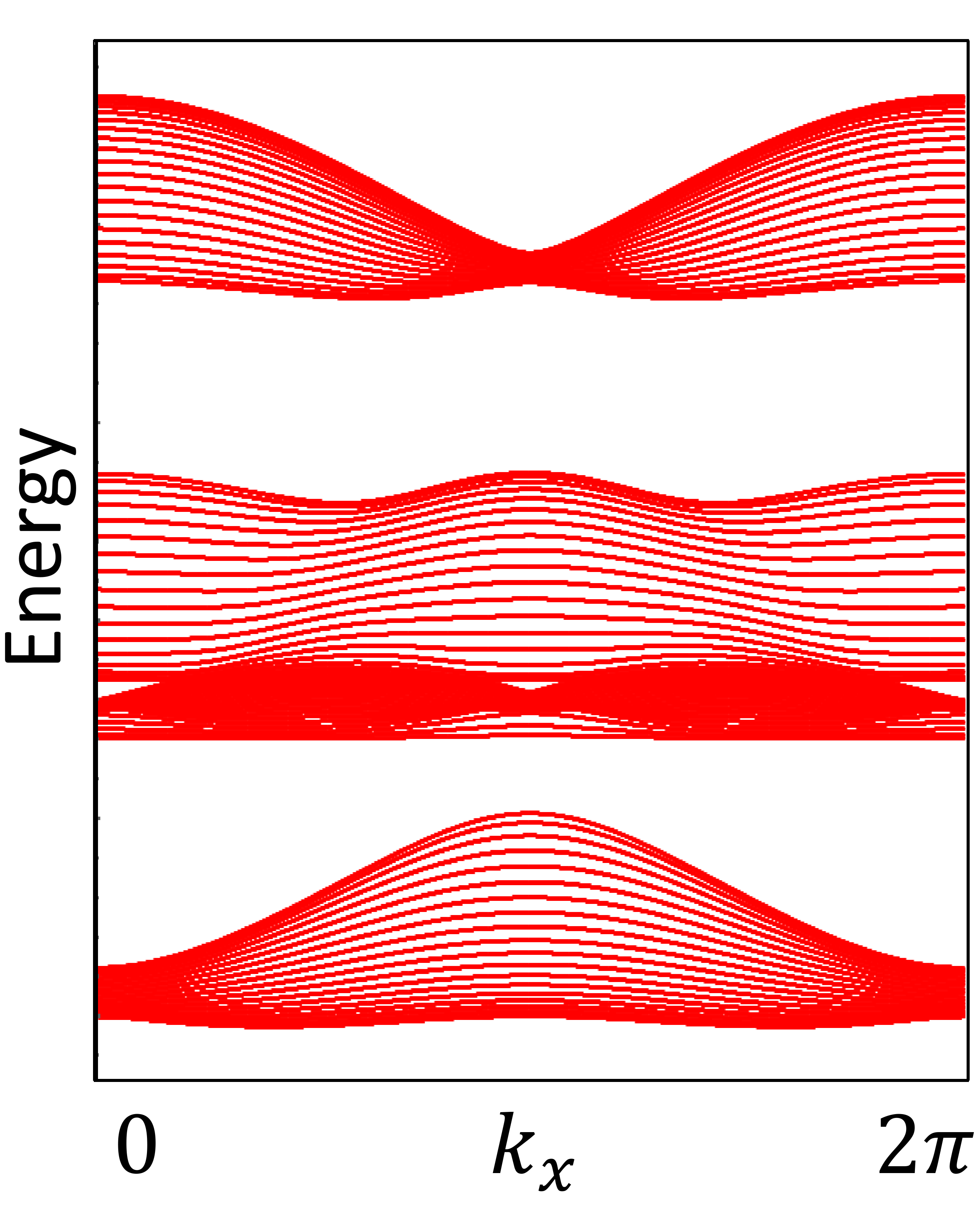}}
\subfloat[]{\label{fig:nontrivial}\includegraphics[width=0.155\textwidth]{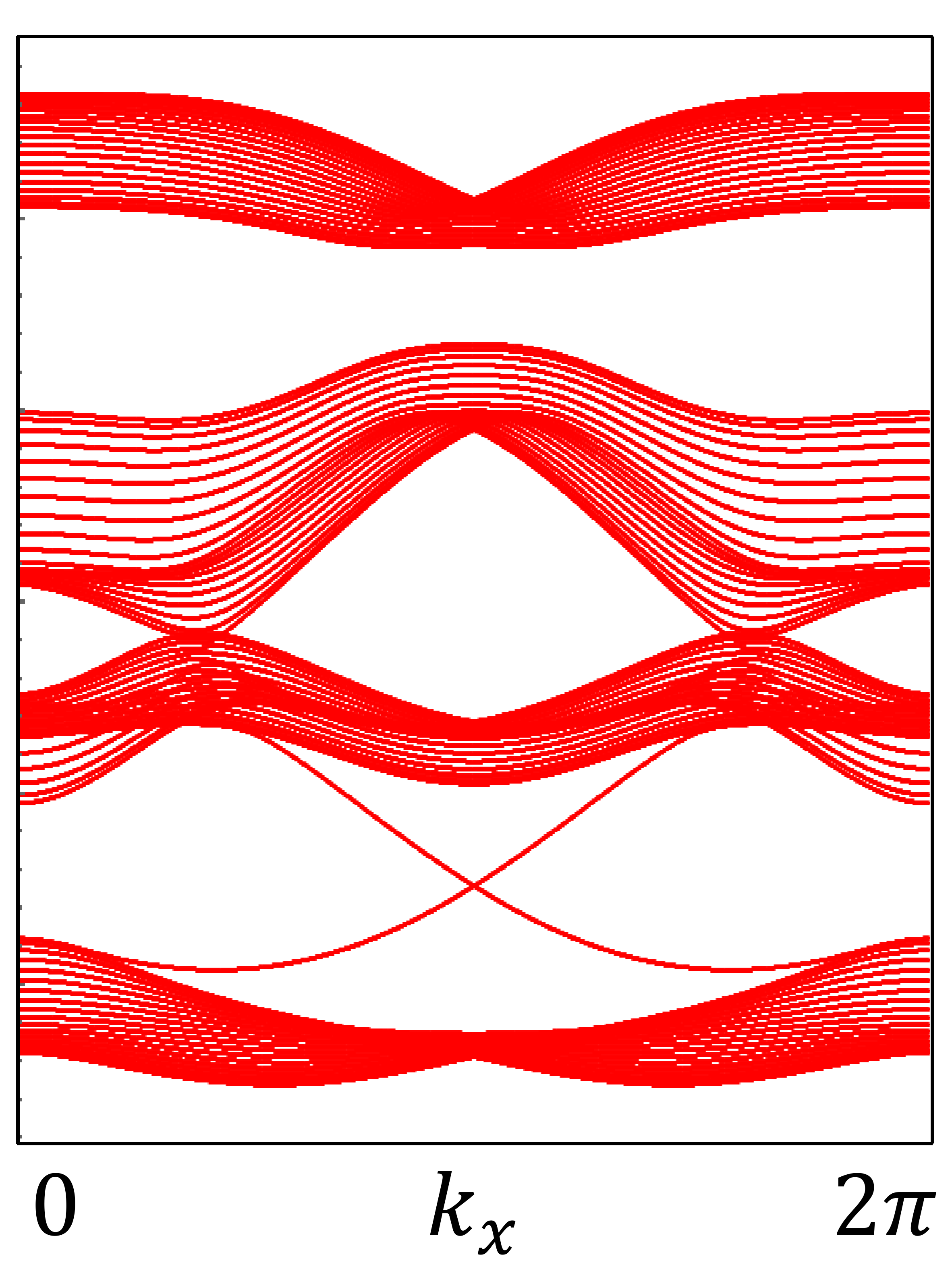}}
\subfloat[]{\label{fig:nontrivial_2}\includegraphics[width=0.155\textwidth]{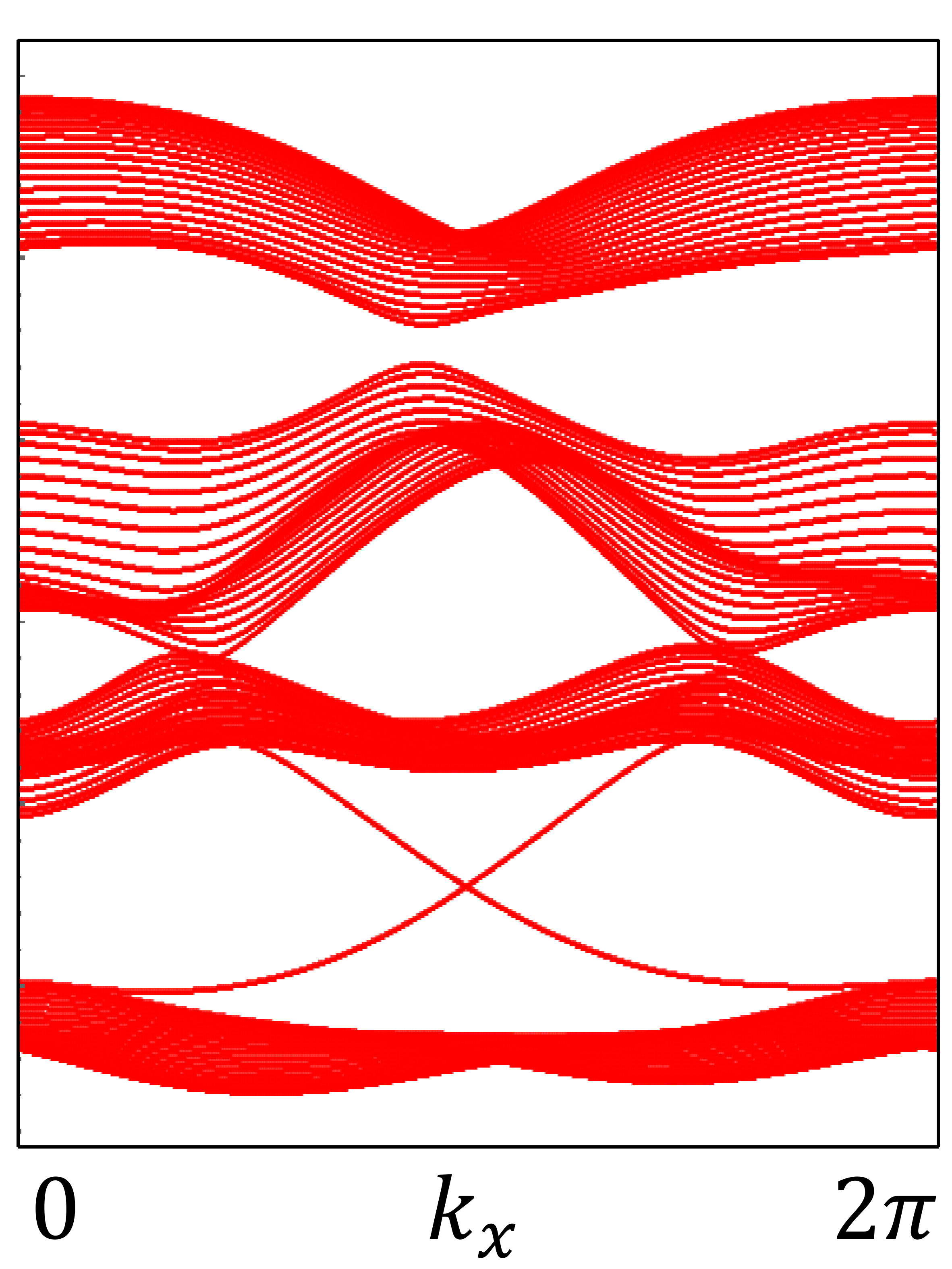}}
\caption{Edge spectrums for trivial and topological insulating phases. (a) Trivial insulator with parameters at empty circle marked in Fig. \ref{fig:phase1}, (b) Topological insulator with parameters at filled circle marked in Fig. \ref{fig:phase1}, (c) Topological phase with parameters which additionally break inversion symmetry from Fig. \ref{fig:nontrivial}. See the main text for more details. }
\label{fig:edge}
\end{figure}
For given parameters, we evaluate the topological invariants which is the product of parity of occupied eigenstates at four different time reversal invariant momentum (TRIM) points in the Brillouin zone. \cite{fu2007topological}  In two dimension, the negative sign indicates a topological phase and the positive sign indicates a trivial phase. In both cases in Eq. \eqref{eq:param} (see also Fig. \ref{fig:distort}), the products of parities at X1 and X2 are related with each other. For case (i) in Eq. \eqref{eq:param}, it satisfies $M^{\dagger}_{\pm\hat{x}+\hat{y}} P M_{\pm\hat{x}+\hat{y}}\!=\!P$, thus the parity at X1 and X2 are the same. For case (ii), however, $C_4^{\dagger}PC_4\!=\!e^{-i(k_x+k_y)}P\!=\!-P$ and the parity at X1 and X2 are opposite.
Thus, one only requires to examine whether the parity at $\Gamma$ and M points are the same or not. As a result, the system at filling $\nu=1$ becomes a topological phase in the following conditions;
\begin{eqnarray}
\text{(i) }&& \sqrt{\tilde{t}_-^2\!+\!u_-^2} \!-\! \sqrt{\tilde{\lambda}^2_-\!+\!u_-^2}\!<\!2u_+\!<\!
\sqrt{\tilde{t}_+^2\!+\!u_-^2}\!-\!\sqrt{\tilde{\lambda}^2_+\!+\!u_-^2}, \nonumber \\
\text{(ii) }&& ~~~~ \tilde{u} ~ \! >\! ~ \tilde{t}_3\!+\!\tilde{t}_4\!-\!|\tilde{\lambda}_+/4 | ~~~ \text{and} \! ~~~\tilde{u} ~ \! >\!  -|\tilde{t}_3\!-\!\tilde{t}_4|\!+\!|\tilde{\lambda}_-/4|
\label{eq:conditions}
\end{eqnarray}
where $\tilde{t}_{\pm}\!=\! 4 (\tilde{t}_1\!\pm\! \tilde{t}_2)$, $u_\pm\!=\!u_1\!\pm u_2$ and $\tilde{\lambda}_{\pm}\!=\! 4 \lambda(1\!\pm\!\delta)$. The conditions are also satisfied when the signs of inequality are all reversed in Eq. \eqref{eq:conditions}.

Based on the conditions in Eq. \eqref{eq:conditions}, possible parameter space for topological and trivial phases are shown in {Fig. {\ref{fig:phase}}}. Figs. {\ref{fig:phase1}} and {\ref{fig:phase2}} show the phase diagrams for case (i) (See also Fig. \ref{fig:distort1}) as functions of $\lambda/\tilde{t}_1$ and $\tilde{t}_2/\tilde{t}_1$ with $u_1\!=\!2$, $u_2\!=\!4$; two different parameter sets (a) for $\delta\!=\!-2$ and (b) for $\delta\!=\!2$ respectively.
Figs. \ref{fig:phase3} and \ref{fig:phase4} represent the phase diagrams for case (ii) (See also Fig. \ref{fig:distort2}) as functions of $\lambda/\tilde{t_3}$ and $\tilde{t_4}/\tilde{t_3}$ with $u_1\!=\!u_2\!=\!1$; (c) for $\delta\!=\!-2$ and (d) for $\delta\!=\!2$ respectively.
In Figs. \ref{fig:phase}(a)-(d), blue color region and white color region are where topological insulating phases and trivial insulating phases are stabilized  if the system is gapped. At the phase boundaries separating two distinct phases, the gap should be closed at least one momentum point to exchange the parity with the unoccupied bands. As shown in Fig. \ref{fig:phase}, there exists large parameter regime for small SOC where topological insulating phases are stabilized with glide reflection symmetry breaking.  

{Fig. {\ref{fig:edge}}} shows the one dimensional band structure in a strip geometry for three distinct cases.\cite{halperin1982quantized,kane2005z, xu2006stability, sheng2005nondissipative, sheng2006quantum} Figs. \ref{fig:trivial} and \ref{fig:nontrivial} show the edge spectrums at parameters marked with empty and filled circles in Fig. \ref{fig:phase1} respectively. One can easily see the absence or presence of edge modes for trivial insulator or topological insulator.
Fig. \ref{fig:nontrivial_2} is the edge spectrum for topological insulator with parameters deviated from the case for Fig. \ref{fig:nontrivial}, in such a way that the system additionally breaks inversion symmetry. More explicitly, we keep all the same parameters except $\tilde{t}_2$ defined in Eq. \eqref{eq:param} but take $t_{11}\!=\!t_{40}\!=\!1.3$, $t_{21}\!=\!2.5$, $t_{30}\!=\!3$  to break inversion symmetry. 
It is worth to note that this topological phase is protected by time-reversal symmetry not by inversion symmetry. These edge states are stable under any small perturbation as long as the perturbation respects time-reversal symmetry. Although the simple analysis of $Z_2$ invariant at TRIM points doesn't work anymore, the topological phase survives even when the inversion symmetry is explicitly broken as shown in Fig. \ref{fig:nontrivial_2}.

{\textbf{\textit { Discussion---}}}
Our analysis discussed so far can be extended beyond the nearest-neighbor tight-binding model and one can still use the indicator Eq. (\ref{eq:conditions}) to explore the parameter regime where topological insulator is stabilized. 
We note that several types of long-range electron hoppings simply enhance the nearest-neighbor hopping parameters for each case discussed in Eq. \eqref{eq:conditions}. In order to see this, let's consider the long-range electron hopping between the sublattices $\alpha$ and $\beta$ with magnitude $\delta t$ across the relative distance ($m, n$) in units of unit cell ($a_x$,$a_y$). Then, the Hamiltonian matrix component in momentum space has additional term $\delta t e^{i (k_x m \!+\! k_y n)}$. 
In both cases (i) and (ii) in Eq. \eqref{eq:param} (see Fig. \ref{fig:distort}), 
any long range hopping connecting A and B sublattices and their symmetry related hoppings results in $\tilde{t}_{1,3}\!\rightarrow\!\tilde{t}_{1,3}\!+\!\delta$ if $m\!+\!n$ is even and $\tilde{t}_{2,4}\!\rightarrow\!\tilde{t}_{2,4}\!+\!\delta$ if $m\!+\!n$ is odd. Here, evenness and oddness have nothing to do with symmetries but are related to the choice of unit cell shown in Fig. \ref{fig:distort}. Similar analysis can be also done for other parameters $u_1$,$u_2$ and $\tilde{u}$.

%


We address another important aspects in our analysis. Through the entire derivation, we have assumed if the system is gapped the topological phases are stabilized with given parameter regime as shown in Eq. \eqref{eq:conditions} and Fig. \ref{fig:phase}. However, analyzing the $Z_2$ invariants in the TRIM points does not always guarantee the insulating phase. This is because there may be accidental gapless points away from the TRIM points in the Brillouin zone. Therefore, it is possible to obtain the negative sign of $Z_2$ invariants even in the absence of the SOC and this originates from the odd number of Dirac cones which are not necessarily at the TRIM points and lead to $\pi$ Berry phases. \cite{fu2007topological, kariyado2013symmetry}   As shown in Fig. \ref{fig:phase}, there exists wide range of parameter space colored in green line where $Z_2$ invariant is negative in the absence of SOC. In this regime, the system must contain the odd number of Dirac cones. 
For case (i) in Eq. \eqref{eq:param}, mirror symmetries guarantee odd number of gapless Dirac points to be along the line between $\Gamma$ and M. In case (ii) in Eq. \eqref{eq:param}, the 4-fold rotational symmetry enforces the gapless point to be located at $\Gamma$ or M point and this indeed happens at $\nu\!=\!1$ and $\nu\!=\!3$ with given parameter set (green line) in Fig. \ref{fig:phase4}. 

Based on our studies, one may expect nonsymmorphic symmetry breaking  could give rise to the phase transition from gapless metal to gapped trivial or topological insulating phases. Hence, the system with strong coupling between lattice and electronic degrees of freedom may favor spontaneous lattice distortion. Due to additional energy gain by opening the gap, the system may  form charge or spin order in such a way that the system breaks nonsymmorphic symmetries. Relevant future work is worth to be explored using the first principle calculation. In addition, one can also imagine possible topological Kondo insulator in nonsymmorphic crystals. In particular, the Kondo lattice model in nonysymmorphic crystals can have interesting behavior in the intermediate Kondo coupling. \cite{pixley2016filling} In this case, the system spontaneously breaks nonsymmorphic symmetry and opens a gap where Kondo insulator is energetically favored. Therefore, depending on how the system breaks nonsymmrophicity, the system can spontaneously drive the phase transition from Kondo semimetal to topological Kondo insulator as a function of Kondo coupling strength. Our results give an insight into the important role of lattice symmetries and their relevance to topological phase transitions and pave the way for exploring relevant materials and experiments in future.

\vskip 1cm


\begin{acknowledgments}

\noindent
{\em Acknowledgments.---}
We would like to thank S. Parameswaran for useful discussions and comments.  H.J.Y and S.B.L. is supported by the KAIST startup and National Research Foundation Grant (NRF-2017R1A2B4008097). 
\end{acknowledgments}

\bibliography{ref}


\end{document}